\documentclass[8.5pt,twoside,twocolumn]{article}
\usepackage[super,sort&compress,comma]{natbib} 
\usepackage[version=3]{mhchem}
\usepackage{times,mathptmx}
\usepackage{sectsty}
\usepackage{balance} 

\usepackage{graphicx} %eps figures can be used instead
\usepackage{amssymb}
\usepackage{marvosym}
\usepackage{lastpage}
\usepackage[format=plain,justification=raggedright,singlelinecheck=false,font=small,labelfont=bf,labelsep=space]{caption} 
\usepackage{fancyhdr}
\usepackage{color}
\begin{document}

\thispagestyle{plain}
\fancypagestyle{plain}{
%\fancyhead[L]{\includegraphics[height=8pt]{headers/LH}}
%\fancyhead[C]{\hspace{-1cm}\includegraphics[height=20pt]{headers/CH}}
%\fancyhead[R]{\includegraphics[height=10pt]{headers/RH}\vspace{-0.2cm}}
\renewcommand{\headrulewidth}{1pt}}
\renewcommand{\thefootnote}{\fnsymbol{footnote}}
\renewcommand\footnoterule{\vspace*{1pt}% 
\hrule width 3.4in height 0.4pt \vspace*{5pt}} 
\setcounter{secnumdepth}{5}

\def\pin{\Pi_{\rm in}}
\def\pout{\Pi_{\rm br}}
\def\peq{\Pi_{\rm eq}}
\def\zetaa{\zeta^*}
\def\ke{\kappa}

\makeatletter 
\def\subsubsection{\@startsection{subsubsection}{3}{10pt}{-1.25ex plus -1ex minus -.1ex}{0ex plus 0ex}{\normalsize\bf}} 
\def\paragraph{\@startsection{paragraph}{4}{10pt}{-1.25ex plus -1ex minus -.1ex}{0ex plus 0ex}{\normalsize\textit}} 
\renewcommand\@biblabel[1]{#1}            
\renewcommand\@makefntext[1]% 
{\noindent\makebox[0pt][r]{\@thefnmark\,}#1}
\makeatother 
\renewcommand{\figurename}{\small{Fig.}~}
\sectionfont{\large}
\subsectionfont{\normalsize} 

\fancyfoot{}
%\fancyfoot[LO,RE]{\vspace{-7pt}\includegraphics[height=9pt]{LF}}
%\fancyfoot[CO]{\vspace{-7.2pt}\hspace{12.2cm}\includegraphics{RF}}
%\fancyfoot[CE]{\vspace{-7.5pt}\hspace{-13.5cm}\includegraphics{RF}}
%\fancyfoot[RO]{\footnotesize{\sffamily{1--\pageref{LastPage} ~\textbar  \hspace{2pt}\thepage}}}
%\fancyfoot[LE]{\footnotesize{\sffamily{\thepage~\textbar\hspace{3.45cm} 1--\pageref{LastPage}}}}
\fancyhead{}
\renewcommand{\headrulewidth}{1pt} 
\renewcommand{\footrulewidth}{1pt}
\setlength{\arrayrulewidth}{1pt}
\setlength{\columnsep}{6.5mm}
\setlength\bibsep{1pt}

\twocolumn[
  \begin{@twocolumnfalse}
\noindent\LARGE{\textbf{Elasticity of compressed microgel suspensions}}
\vspace{0.6cm}

\noindent\large{\textbf{Giovanni Romeo${\dagger}$\textit{$^{a,b}$} and Massimo Pica Ciamarra\textit{$^{c}$} }}\vspace{0.5cm}

%\noindent\textit{\small{\textbf{Received Xth XXXXXXXXXX 20XX, Accepted Xth XXXXXXXXX 20XX\newline
%First published on the web Xth XXXXXXXXXX 200X}}}

\noindent \textbf{\small{DOI: 10.1039/c3sm50222h}}
\vspace{0.6cm}
%Please do not change this text.

\noindent \normalsize{
We describe the elasticity of compressed microgel suspensions 
starting from a core--shell model for the single particle.
The mechanical properties of the inner core and of the outer corona are 
described via the mean field Flory theory, and via the Alexander--de Gennes
model for polymer brushes, respectively. The model successfully reproduces experimental measures of the
elastic shear modulus up to a constant factor that we rationalize in the jamming perspective.
}
\vspace{0.5cm}
 \end{@twocolumnfalse}  ]

\section{Introduction}
\footnotetext{\textit{$^{a}$~Italian Institute of Technology@CRIB, Napoli, Italy}}
\footnotetext{\textit{$^{b}$~Department of Materials Engineering and Production, University of Napoli Federico II, Italy}}
\footnotetext{\textit{$^{c}$~CNR-SPIN, Dipartimento di Scienze Fisiche, University of Napoli Federico II, Italy. E-mail: massimo.picaciamarra@na.infn.it}}
\footnotetext{\textit{${\dagger}$~While this manuscript was finalized, the leading author, Dr. Giovanni Romeo, 
passed away unexpectedly. I wish to express my deepest sorrow for the loss of a dear friend and passionate researcher. Giovanni will be deeply missed by many
and warmly remembered by all.}}
Microgels are colloidal particles made of a polymer gel swollen in a solvent.
They are extensively investigated due to their role in applications including purification technology and sensing~\cite{Das2006,Guo},
and are used to modify the viscoelastic properties of soft matter systems for
drug delivery and cosmetics~\cite{Peppas94,Nolan05}.
Microgel suspensions are useful in this context because
they can percolate the whole space at very low polymer concentration,
thus influencing the viscoelastic properties of the medium.
Moreover, these properties are easily tunable as they
smoothly increase with particle concentration. This occurs because particles
are soft, and in dense suspensions they shrink and deform.
The dependence of the particle size on concentration is peculiar to these soft suspensions, 
and it is not observed neither in atomic nor in the majority of colloidal systems.
This peculiarity makes difficult to understand the dependence
of their viscoelastic properties on concentration, 
which is commonly expressed in terms of a
generalized volume fraction $\zeta = n · v^0$~\cite{micbook, Romeo10, Mattsson09, Lyon12}. Here $n$ is the number density
and $v^0$ the volume of the particle measured in the dilute limit, $n \to 0$.
Recent experimental results~\cite{Lietor11} suggest that the mechanical properties of these dense suspensions
are only determined by the single particle bulk modulus $k_p$. 
For instance, it is found that the shear modulus $G'$ and the osmotic pressure $\Pi$ scale with $k_p$,
as the ratios $G'/k_p \ll 1$ and $\Pi/k_p~\simeq 1$ are $\zeta$ independent~\cite{Lietor11}. 
These results indicate that, since the osmotic pressure sets the bulk modulus $k_p = \zeta d\Pi/d\zeta$, 
the key to the concentration dependence of the elastic properties of concentrated microgel suspensions, is the $\zeta$
dependence of $\Pi$.
Indeed, modeling a microgel particle as an hard core surrounded by a corona of polymer brushes~\cite{Alexander}, Scheffold and
co-workers were able to estimate $\Pi(\zeta)$~\cite{Scheffold10} and, from this, to predict the shear modulus $G'(\zeta)$ up to a constant $a_g = G'/\Pi \ll 1$. 
This estimate qualitatively reproduces experimental data close to $\zeta = \zeta_J \simeq 1$~\cite{Scheffold10}, but predicts
a divergence of $G'$ at higher $\zeta$ as a consequence of the assumed core incompressibility.
This is in contrast with experimental results in the highly packed regimes, $\zeta \gg 1$, 
showing that the elasticity remains finite, and is qualitatively described in the framework of macroscopic polymer gel theories~\cite{Flory,Menut12}.
Currently, a single model able to describe the concentration dependence of the osmotic pressure and of the shear modulus in a 
concentration range spanning from the jamming point up to highly packed and compressed states is lacking. 
In addition, the physical origin of the small scaling factor $a_g$ is unknown.

Here we show, via an experimental and theoretical study, that the shear elasticity of 
microgel suspensions can be described in terms of a model accounting for the particle deformation, 
which allows to estimate the concentration dependence of the osmotic pressure.
The model assumes a particle as composed by a corona of polymer brushes
attached to a compressible core, and we describe their osmotic pressures
via the Alexander--de Gennes model~\cite{Scheffold10}
and the Flory--Rehner mean--field theory~\cite{Flory}, respectively.
At every concentration the relative compression of the particle core and corona is related to
an equilibrium osmotic pressure which sets the suspension elasticity $G'$. 
The model captures well the $\zeta$ dependence of  $G'$, and 
confirms that $G'$ is orders of magnitude smaller than $\Pi$,
$a_g \ll 1$. We suggest that the small value of $a_g$
can be related to the non--affine response of the dense suspensions, 
and we rationalize its origin in the framework of the jamming scenario~\cite{vanHecke}.

\section{Experimental}
We have validated our theoretical model against three different set of microgel
particles. 
Here we describe their synthesis processes,
their geometric structure and their physical features. 
In addition, we detail the experimental measure of the osmotic pressure.

\subsection{Particle synthesis \label{par:synthesis}}
We prepare three different sets of microgel particles.
Two sets are obtained by co-polymerization of
poly-N-isopropylacrylamide (pNipa), a neutral polymer, and acrylic acid which adds ionic groups to
the polymer network, with a cross--linker concentration of $0.5\%$. 
Since these microgels respond to temperature and pH variations,
two different sets are obtained by changing the solution pH, at $T = 25^{\circ}$C.
When $7\leqslant \textrm{pH} \leqslant 8$, the ionic groups are completely dissociated,
resulting in fully swollen microgel particles, we will refer to as ionic particles.
When $\textrm{pH} \simeq 3$, the ionic contribution is negligible and the swelling
ratio depends on temperature. We will refer to these particles as neutral particles.
A third set of particles is prepared by polymerization of pNipa with a cross--linker concentration of $3\%$.
The size of these highly cross--linked particles is pH independent, and we investigate their properties
in the fully swollen state at $T = 10^{\circ}$C.

\subsection{Characterization of the particle structure \label{par:structure}}
\begin{figure}[t!]	
\includegraphics*[scale=0.52]{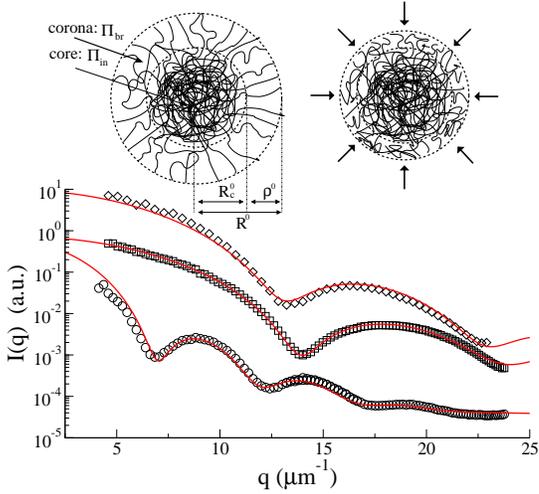}
\caption{\label{fig:model} (color online) Light scattering
measurements of highly cross--linked (diamonds), at $T = 10^{\circ}$C,
and of neutral (squares) and of ionic (circles)
microgel suspensions, at $T = 25^{\circ}$C. Lines are the corresponding fit to the form factor described in the text.
The fits allow to estimate the size of the core $R_c^0$ and of the corona $\rho^0 = 4\lambda$ of a swollen particle. 
The drawings schematically illustrate a swollen (left) and a compressed (right) microgel.}
\label{structure}
\end{figure}

Since the compression of a microgel depends on its structure, we estimate the radial
polymer density distribution within a particle via scattering experiments. 
It is well established that a microgel is characterized by a core with uniform polymer density, and by
a corona in which the density decreases towards the particle
periphery~\cite{Stieger04, Scheffold09, Mason05}, as schematically illustrated
in Fig.~\ref{fig:model}.

The core and the corona size can be measured by fitting the wavevector, $q$,
dependence of the light intensity $I(q)$ scattered  by dilute suspensions, to an
inhomogeneous sphere form factor~\cite{Stieger04} like:
$\widetilde{P}(q)\propto
\left[3\frac{Rq\cos(Rq)-\sin(Rq)}{(Rq)^3}\cdot\exp\left(\frac{-\lambda^2
q^2}{2}\right)\right]^2$.
Here the form factor of an homogeneous sphere of radius $R$ is convoluted
with an exponential factor accounting for the density decay over the
characteristic length scale $\lambda$.
To account for polydispersity, we assume a 
Gaussian distribution $W(R, \overline{R}, \sigma)$
for the radius $R$, with average value $\overline{R}$ and variance $\sigma^2$.
The form factor for dilute suspensions is thus described by $P(q)= \int
\widetilde{P}(q)\cdot W(\overline{R}, R, \sigma)\ dR$.

The quality of the fit of the form factor of our suspensions, shown in
Fig.~\ref{fig:model}, supports the core--shell model.
We find $\overline{R} = 0.4\mu$m, $\lambda \simeq 0$ and $\sigma= 0.04$ for the neutral particles,
$\overline{R} = 0.65\mu$m, $\lambda = 0.06\mu$m and $\sigma= 0.04$ for the ionic ones, and
$\overline{R} = 0.35\mu$m, $\lambda = 0.01\mu$m and $\sigma= 0.03$ for the highly cross--linked particles.
The overall particle radius $R^0$ and the core radius 
$R^0_c$ are related to these values~\cite{Stieger04} by
$R^0 = \overline{R}+2\lambda$ and by $R^0_c = \overline{R}-2\lambda$,
respectively.

\subsection{Thermodynamic properties of the microgel \label{par:thermo}}
Using the Flory-Rehner's theory for polymer gels~\cite{Flory}, the mixing
contribution to the free energy is given by
\begin{equation}
\pi_m/kT=
-\frac{N_a}{v_s}\Big[\frac{\varphi_0}{\alpha}+\ln(1-\frac{\varphi_0}{\alpha}
)+\chi\big(\frac{\varphi_0}{\alpha}\big)^2\Big],
%\label{pm}
\nonumber
\end{equation}
while that due to the elasticity of the polymer chains is
\begin{align}
\pi_{el}/kT=\frac{n_c}{v_0}\Big[\frac{\varphi_0}{2\alpha}-\frac{1}{\alpha^{1/3}}
\Big].
%\label{pe}
\nonumber
\end{align}
Here $kT$ is the thermal energy,
$\alpha=v/v_0=\varphi_0/\varphi$ is the volumetric
swelling ratio, $N_a$ is the Avogadro's number, $v_s$ the
molar volume of the solvent and $n_c$ the effective number of chains
in a microgel particle, $\chi$ the polymer--solvent interaction parameter. 
In addition, $\varphi$ and $v$ are the
polymer volume fraction and the particle volume, respectively, and
$\varphi_0$ and $v_0$ are their values in the microgel generation
state, the deswollen state in our case. 

For the particles at cross--linker concentration of $0.5\%$, we use 
the experimental measurement of the swelling ratio $(R/R^0)^3 =33$ in
dilute solution at temperature $T = 7 \ ^oC$  and pH $\simeq 3$, and the condition that $\pin=0$ with
$\chi=0$, $v_s=18$cm$^3$, $v_0= 0.0115\mu$m$^3$, $\varphi_0=0.7$, to determine $n_c =
2.9 \times 10^5$. We observe that for $T\lesssim 20 \ ^oC$ water behaves like an athermal
solvent for Nipam justifying the assumption $\chi=0$ \cite{Crassous08}.
For the particles at cross--linker concentration of $3\%$, we use 
the same procedure as before to determine $n_c = 1.05 \times 10^7$.

In ionic systems there is an additional osmotic
pressure $\pi_i$ arising from the presence of charged groups in the polymer
network. In the absence of salt, the main contribution to this ionic
term is given by the gas of counterions within the particle.
Assuming electroneutrality,
the number of counterions inside a particle is equal to the total
number of ionized groups per particle. As a result:
\begin{equation}
\pi_i/kT=\frac{Q}{v_0\alpha},
%\label{pi_i}
\nonumber
\end{equation}
where $Q$ is the number of ions per particle.
For our particles, $n_c = 2.9 \times 10^5$ and $Q=7 \times 10^6$ \cite{Romeo10},
and Flory theory predicts $(R/R^0)^3 =140$. From dynamic light scattering
we find that at $pH=7$ and $T=20 \ ^oC$  $R=0.8\mu$m which leads to
$(R/R^0)^3 =180$. The agreement between the prediction and the experiments is
reasonable, considering that Flory theory does not account for the polymer density
inhomogeneity in the microgels.

\subsection{Measure of shear modulus and osmotic pressure \label{par:osmotic}}
We measure the elastic shear plateau modulus $G'(\zeta)$ of the suspensions 
via linear frequency sweep scans, using rheometers (ARG2, TA Instruments and MCR302, Anton Paar) with plate--plate geometries.
As common in suspensions of soft microgel particles, 
a plateau modulus emerges at volume fractions $\zeta > \zeta_J \simeq 1$.
This value is larger than the jamming threshold for stiff particles, $\zeta \simeq 0.64$, possibly because
particles deform at constant volume in the volume fraction range $0.64$--$1$, without an appreciable energy cost.

To directly measure the osmotic pressure of dense suspensions, a microgel
solution at $\zeta_0 \gg \zeta_j$ and  
$pH \simeq 3$ is transferred into bags made from dialysis membranes (SpectraPor
regenerated cellulose membrane, MWCO 12--14 kDa) 
and immersed in water solutions of dextrans (Fluka, Mn=70 KDa) at different
concentrations. 
The relationship between the dextran concentration and the resultant osmotic
pressure has been calibrated previously~\cite{Cabane2004}.
After equilibration for one week we measure the weight $w$ of the solution
contained in the bag; from this the actual microgel concentration is determined
as:  $\zeta=\zeta_0(w_0/w)$ where $w_0$ is the initial weight at concentration
$\zeta_0$.

%The estimates of $R^0$ are in good agreement with the hydrodynamic radius
%$R_h$ measured by dynamic light scattering, $R_h=0.45\mu$m and $R_h=0.8\mu$m for the neutral and ionic particles, respectively.

\section{Theoretical model}
The microgel structure suggests that, from a mechanical point of view, a
particle can be described by a series of two non--linear springs of length
$\rho = R-R_c$ and $R_c$, where $R$ is the radius of the overall particle, 
and $R_c$ the radius of the homogeneous core region.
When concentration is increased, particles compress and the
resulting deformation of the two springs is accounted by 
the relative variation of $R_c$ and $\rho$ with respect to their values in
dilute solutions, $R_c^0$, and $\rho^0 = R^0-R_c^0$, for fixed external
conditions (temperature, pH).

\subsection{Corona: Alexander--de Gennes model}
The outer spring originates from the interaction between the outer polymer chains, which 
results in a pressure $\pout$ described by the Alexander--de Gennes model~\cite{Alexander}. 
This model relates the pressure to the length $\rho$ of the polymer brushes
which make the outer corona~\cite{Scheffold10}
\begin{equation}
\pout(\rho)=\frac{kT}{s^3}\left[
\left(\frac{\rho}{\rho^0}\right)^{\nu_{\rm
os}}-\left(\frac{\rho}{\rho^0}\right)^{\nu_{\rm el}}
\right],
\label{eq:adg}
\end{equation}
where $s$ is the mean distance between the anchoring points of the surface
polymer chains. In the corona region,
and within the assumption of an isotropic cross-link distribution, this must be similar to the characteristic decay length, 
and we will therefore assume $s = \lambda$ in the following.
%where $s$ is the mean distance between the anchoring points of the surface polymer chains. 
%In the corona region, this is similar to the length of
%the polymer brushes, and therefore we set $s = \lambda$.
The first term on the right hand side of Eq.~\ref{eq:adg}
is the osmotic pressure of a semidilute polymer solution and the second is the
pressure related to chain elasticity. 
The exponents are fixed by the dependence of the characteristic blob size $\xi$, in the semidilute regime, on polymer concentration. 
Indeed, the osmotic pressure $\pi$ and chain elastic force $f$ can be calculated as $\pi=kT/\xi^3$ and $f=kT/\xi$~\cite{Rubinstein03}.
For neutral particles, $\nu_{\rm os} =
-9/4$ and $\nu_{\rm el} = 3/4$,
while for ionic ones, $\nu_{\rm os} = -3/2$ and $\nu_{\rm el} =
1/2$~\cite{Rubinstein05}.
In the case of ionic particles, one should also account for the contribution to the osmotic pressure due to counterions in the corona regions.
When the electrostatic screening length is much smaller than
$\rho^0$, as for our particles~\cite{Romeo12}, this contribution is relevant when
$\rho < \rho^0$, and a term
$kT (Q/n_cs^2) (1/\rho^0-1/\rho)$~\cite{Pincus1991} has to be added to
Eq.~\ref{eq:adg}.
Here $Q$ and $n_c$ are the number of ions and the number of polymer chains per
particle; $Q/n_c$ is the number of counterions per chain.

\subsection{Core: Flory--Rehner mean--field theory}
The inner spring models the compression of the core,
a swollen polymer network. The osmotic pressure $\pin$ of this network
is described by the Flory--Rehner mean--field theory~\cite{Flory},
which relates $\pin$ to the core size,
\begin{equation}
\pin(R_c)=\pi_m(R_c)+\pi_{el}(R_c).
\label{eq:pin}
\end{equation} 
In this equation, the first term, $\pi_m$, accounts for the
variation in free energy due to polymer/solvent mixing. The second
term, $\pi_{el}$, is the elastic contribution resulting from the deformation of the polymer chains
with respect to their equilibrium configuration.
In the case of ionic particles, the inner pressure $\pin$ also includes a term
$\pi_i(R_c)$ describing
the additional osmotic pressure arising from the presence of a gas of
counterions within the polymer network. We remark that the systems studied here are salt-free. However salt effects can be easily included in the expressions of $\pin$ and $\pout$ with standard approaches \cite{Rubinstein05, Yigit12}. 
The Flory--Rehner~\cite{Flory} mean--field theory gives the dependence
of the different terms of $\pi_m$, $\pi_{el}$ and $\pi_i$ on the particle and
solvent properties as described in the previous section. 

\subsection{Equilibrium pressure}
The condition of mechanical equilibrium between core and corona,
$\pin = \pout= \peq$, together with the condition that above $\zeta_J$ the
particle size $R_c+\rho$ decreases as \cite{Romeo12}
\begin{equation}
\frac{R_c+\rho}{R_c^0+\rho^0} = \left(\frac{\zeta}{\zeta_J}\right)^{1/3},
\label{eq:size}
\end{equation}
allow to determine the equilibrium pressure $\peq$, and the relative compression
of the core $R_c/R_c^0$ and of the corona $\rho/\rho^0$ at each value of the volume fraction
$\zeta$. 
The equilibrium pressure $\peq$ sets the microscopic elastic constant $k$, which in the Derjaguin
approximation~\cite{Israelachvili} is $k = \pi R_c \peq$ \cite{Scheffold10}. This allows to estimate
the macroscopic shear modulus~\cite{Scheffold10, Cloitre03, Zwanzig} as 
\begin{equation}
G'_{\rm th} = a_g k /\pi R = a_g  \frac{R_c(\zeta)}{R(\zeta)} \peq(\zeta).
\label{eq:Gteo}
\end{equation}
Eq.~\ref{eq:Gteo} neglects the thermal contribution to the shear elasticity, which is
the only contribution present for $\zeta < \zeta_J$. 
This contribution is of the order of tens of mPa, and can be safely neglected as it is not usually experimentally detected.

\section{Results}

\subsection{Concentration dependence of the shear elasticity}
In the limit of negligible core compression, our model reduces to the Alexander-De Gennes polymer brush
interaction~\cite{Alexander,Scheffold10}. In this approximation
the prediction correctly describes the elastic shear modulus of microgels with a
stiff core at volume fractions $\zeta \gtrsim  \zeta_J$, up to a constant scaling factor $a_g$~\cite{Scheffold10}.
\begin{figure}
\includegraphics*[scale=0.33]{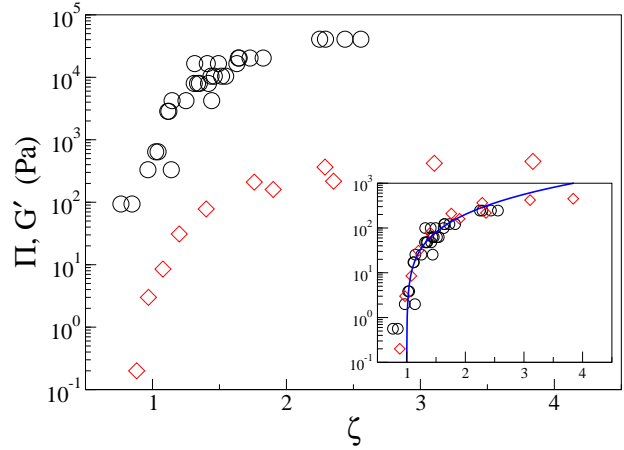}
\caption{\label{fig:nonionic} 
(color online) Concentration dependence of the osmotic pressure $\Pi$ (circles)
and of the shear
modulus $G'$ (diamonds) of neutral microgel suspensions.
Inset: Experimental $\Pi$ and model prediction (full
line), vertically rescaled on the experimental $G'$.}
\end{figure}
The opposite limit of negligible corona compression is expected to describe our neutral suspensions,
since the corona of neutral particles has a negligible length, $\lambda \simeq 0$. In this case, 
Eq.~\ref{eq:size} simplifies to $R = R_c = R_c^0
\left(\zeta/\zeta_J\right)^{1/3}$, 
and Eq.~\ref{eq:Gteo} to $G' = a_g \peq$, where the osmotic pressure $\peq =
\pin$ is readily estimated from Eq.~\ref{eq:pin}.
To validate this prediction, we have directly measured both the osmotic pressure
$\Pi$ as well as the shear modulus $G'$ of the neutral suspensions at different
concentrations. %~\cite{supplementary}.
The experimental data in Fig.~\ref{fig:nonionic} show that $G'$ is orders of
magnitude smaller than $\Pi$, consistently with previous results~\cite{Lietor11}.
However $\Pi$ and $G'$ share the same concentration dependence, and such
dependence is captured by the model. 
Indeed the inset of Fig.~\ref{fig:nonionic} shows that the
the model prediction describes the experimental $G'$ when rescaled by a factor
$a_g \simeq 5 \cdot 10^{-3}$. The experimentally measured pressure $\Pi$ also collapses
on $G'$, when rescaled by $\simeq a_g$.
This result strongly supports the idea that the shear elasticity of the
suspension is intimately related to the osmotic pressure~\cite{Lietor11}, 
and shows that the model correctly describes $\Pi(\zeta)$.
We note that the intimate relation between the shear elasticity and the
osmotic pressure also exists for other disordered soft materials like emulsions~\cite{Mason95}.

\begin{figure}[!t]
\includegraphics*[scale=0.33]{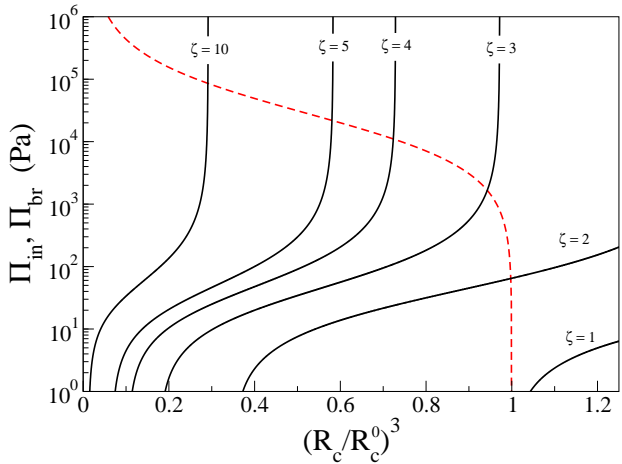}
\caption{\label{fig:pressioni}
(color online) Dependence of the inner pressure $\pin$ (dashed line),
and of the brush pressure $\pout$ (solid lines) of ionic suspensions, on the relative core volume. $\pout$ is estimated at different $\zeta$ values, as indicated.}
\end{figure}

\begin{figure}[!t]
\includegraphics*[scale=0.33]{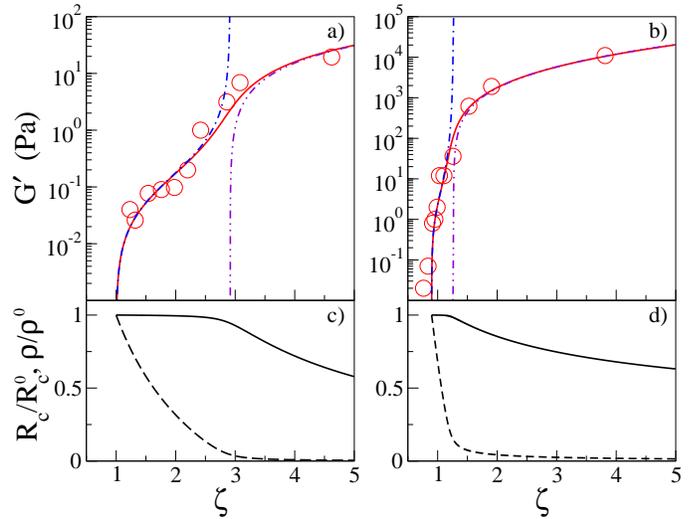}
\caption{\label{fig:ionico} 
(color online) Experimental shear modulus (circles) and
model prediction (full line) of ionic (a) and
of highly cross-linked (b) microgels.
The predictions in the limit of incompressible core 
(dash-dotted line) and of no corona (dash-double dotted line) are also reported.
Panels c) and d) illustrate the corresponding model predictions for the relative compression of the core (full line) and
of the corona (dashed line).}
\end{figure}

When neither the compression of the corona nor that of the core can be
neglected, as for our ionic particles, the full model must be employed to describe the
elasticity of the suspension.
In this case the size of the core and of the corona are fixed by
Eq.~\ref{eq:size} and by the condition of mechanical equilibrium $\pin =
\pout$. 
For each value of $\zeta$, we numerically solve these equations comparing
$\pin$ and $\pout$, as illustrated in Fig.~\ref{fig:pressioni}.
The inner pressure $\pin$, only depends on the core size $R_c$.
$\pin = 0$ in absence of core compression, $R_c = R_c^0$, and
monotonically increases with decreasing $R_c$.
Conversely, the brush pressure $\pout$ monotonically increases with $R_c$.
Indeed, as $R_c$ increases, $\rho$ decreases and varies between the limits $\rho =
\rho^0$ and $\rho \to 0$, where the pressure $\pout$ vanishes and diverges,
respectively. 
For every value of $\zeta$, the monotonic dependence of $\pout$ and $\pin$ on
$R_c$ assures the existence of a single equilibrium pressure $\peq = \pin = \pout$,
which fixes $R_c$ and $\rho$.
From $\peq(\zeta)$ we estimate the $\zeta$ dependence of the elastic shear modulus $G'_{\rm th}$ using Eq.~\ref{eq:Gteo}.
\noindent The theoretical prediction well describes the experimentally measured $G'$
when rescaled by a factor $a_g \simeq 5 \cdot 10^{-3}$, as illustrated in Fig.~\ref{fig:ionico}a.
The full model must also be employed to describe the suspensions of highly cross--linked 
particles, whose corona is not negligible. The experimental $G'$ is illustrated 
in Fig.~\ref{fig:ionico}b, along with the model prediction rescaled by a factor $a_g=G'/ \Pi \simeq 5\cdot10^{-3}$.
Differently from what observed in absence of the core--shell structure, Fig.~\ref{fig:nonionic},
in these cases the shear elasticity $G'(\zeta)$ grows with different regimes. 
A shear elasticity growing in a similar manner has also been
observed in other microgel systems~\cite{Cloitre03a,Lyon12,Menut12}.
%The inset shows the model predictions for the relative compression of the core and of the corona.
%Also in this case a qualitative change in the $\zeta$ dependence of $G'$ occurs when the brush is fully compressed.
This behavior, originates from the $\zeta$ dependence
of the relative compression of the core, $R_c/R_c^0$, and of the corona,
$\rho/\rho_0$.
Indeed, Fig.s~\ref{fig:ionico}c,d reveal that as the volume fraction increases
the corona compresses before the core.
This indicates that the core is stiffer than the corona, consistently with
its higher polymer concentration. The concentration at which the core
starts compressing roughly corresponds to that at which the $\zeta$ dependence of $G'$ changes.
This suggests that the small and the high $\zeta$ behaviour of $G'$
result from the compression of the corona and of the core, which
can be approximated by the incompressible--core and the no--corona limits of
our model.
Indeed, Fig.s~\ref{fig:ionico}a,b show that the shear elastic modulus computed using
these limits correctly captures the different $G'$ regimes.
The model clarifies that the separation between these two regimes depends on the
particle structure, which explains why they are not always clearly distinguished
in the rheological data.
\begin{figure}[!t]
\includegraphics*[scale=1.3]{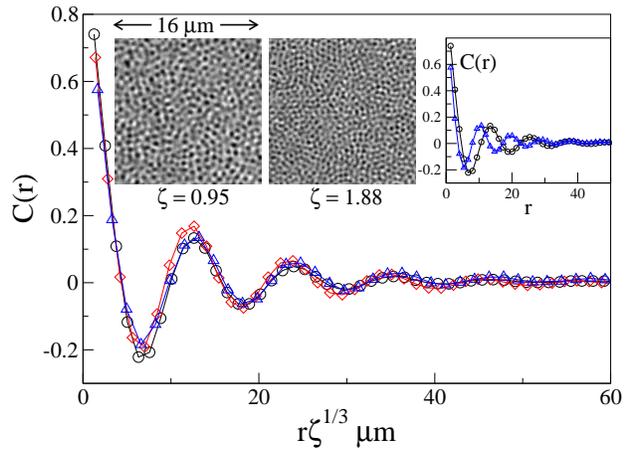}
\caption{\label{fig:pixel} 
(color online) 
Collapse of the radially averaged pixel-pixel correlation function of microscopy images
of the suspensions of highly cross-linked microgels, with concentrations $\zeta = 0.95$ (black circles), $\zeta = 1.16$ (red diamonds) and
and $\zeta = 1.88$ (blue triangles). The inset shows two microscopy images at different volume fractions, and
the relative unscaled correlation functions.}
\end{figure}

%\subsection{Anomalous ratio between shear and bulk modulus of soft particle disordered suspensions}
\subsection{Anomalous ratio between shear and bulk modulus}
To rationalize the concentration independence of the scaling factor $a_g$, and its
small value, we consider that the elasticity of solids is conveniently
described via spring-network models~\cite{Thorpe1999}. In this picture, the nodes represent the
particles, and the springs their force of interaction, in the linear approximation.
The moduli depend on the average stiffness of the springs, $k$, or equivalently
on the average single particle bulk modulus, and on the geometrical
features of the network. Precisely, under the assumption that the geometrical features of the network are fixed by the
density, $\phi$, the bulk and the shear modulus are expressed as $K = k(\phi) f_K(\phi)$, and as $G = k(\phi) f_G(\phi)$, respectively.
The functions $f_K$ and $f_G$ summarize the role of the structure on
the moduli. They are generally unknown, but can be determined in the case of crystals, 
and therefore of ordered networks, by exploiting 
periodicity to concisely describe the structure~\cite{Ashcroft}. 
This picture clarifies that, in general, density affects the elasticity
influencing both the single particle bulk modulus, as well as
the geometrical structure. For instance, in the jamming perspective~\cite{vanHecke},
above the jamming threshold signaling the onset of mechanical
rigidity, one finds $f_K(\phi) = 1$, which is the value expected in the
presence of an affine deformation of the network on compression,
and $f_G(\phi) \propto (\phi-\phi_J)^{1/2}$.
%Close to the transition, the ratio between the moduli vanishes as $G/K \propto (\phi-\phi_J)^{1/2}$.
The small value of the shear modulus close the transition originates from
the presence of a strong non-affine response of the system. This means
that under an applied external shear, the single particle motion does
not simply follow the imposed external deformation. Indeed, the presence of non-affinity
enhances the fluctuation term of the stress tensor, which gives a negative contribution
to the moduli~\cite{fluctuation_modulus}.
In the case of microgel suspensions, $f_K(\zeta) = 1$ as
the suspension bulk modulus equals the single particle bulk modulus~\cite{Lietor11}, 
which is fixed by the osmotic pressure. As a consequence, $a_g = f_G(\zeta)$,
and its constant value thus suggests that 
the structure of the system is concentration independent.
This is speculation is in agreement with recent results obtained with our ionic particles~\cite{Romeo12},
that show that the particle size varies with concentration as $\zeta^{-1/3}$ in dense systems.
%We rationalize the concentration independence of the scaling factor $a_g$ by assuming that
%the shear modulus can be expressed as 
%$G' \sim K f(\vec r)$, where $K$ is the bulk modulus, set by the single particle properties, and 
%$f(\vec r)$ a function characterizing the structure of the packing, with $\vec r$ indicating the ensamble of the particle positions.
%Usually, the concentration affects the structural properties and therefore $f(\vec r)$ is not constant.
%For example, in the jamming framework~\cite{vanHecke}, $f(\vec r)$ 
%is related to the excess contact number $\Delta Z = Z-Z_J$, which increases with particle concentration; 
%here $Z$ is the mean number of contacts per particle, and $Z_J$ its value at the jamming transition. %, $Z_J = 6$ in three dimensions.
%In microgel suspensions $K\simeq \Pi$ \cite{Lietor11}, and therefore $a_g \simeq f(\vec r)$. 
%Accordingly, $a_g$ changes with $\zeta$ if the structure of the suspension changes with $\zeta$. 
%However recent results obtained with our ionic particles \cite{Romeo12} show that the particle size varies with concentration as $\zeta^{-1/3}$, 
%which suggests that the structure might be concentration independent. 

To verify this scenario, we focus on the suspensions of high cross--linked microgels.
Indeed, the structure of these suspensions can be directly investigated as the
high cross--linking density makes these particles visible via optical microscopy.
Images of the suspension at three different concentrations are shown in Fig.~\ref{fig:pixel} (inset). 
To obtain a structural information from the 2D images, we calculate their radially averaged pixel-pixel correlation function $C(r)$.
Fig.~\ref{fig:pixel} shows that the correlation functions collapse
on a single master curve when distances are rescaled by $\zeta^{-1/3}$,
confirming that the particle size decreases as $\zeta^{-1/3}$~\cite{Romeo12}.
In addition, the collapse of the $C(r)$ demonstrates that the geometrical features of the packings
are concentration independent, and that $a_g$ is fixed by 
structure of the packing. 
As a side remark, we note that this scenario might also be relevant for emulsions,
that also have a constant $G/K$ ratio~\cite{emulsions}. In this case, the geometrical
properties of the contact network might be concentration independent, as above jamming
particles deform on increasing concentration, being incompressible.

The presence of a concentration independent structure explains the concentration
independence of $a_g = G/K$, not its small value. This must arise from a 
strong non-affine response~\cite{fluctuation_modulus} of the microgel suspensions,
which is concentration independent being the structure concentration independent.
Non-affinity in suspensions of compressible particles has not been investigated
so far, and we hope that our results stimulates future work in this direction. 
In this respect we note that a single microgel particle, being a gel network,
is expected to deform non-affinely~\cite{Basu2012}. Accordingly, 
one needs to characterize the non-affine behavior
of a suspension of non-affine particles.

\section{Conclusions}
In this work we have described the elastic properties of microgel suspensions
on the basis of a simple model for the particle structure and mechanical properties.
Such model effectively links the sub--particle scale to the macroscopic one,
highlighting the polymer--colloid duality of microgel particles.
Our results show that 
the elastic properties of microgel suspensions
reflect the presence of two characteristic length scales in the particle structure.
In addition, our results clarify that the unexpected relation between the bulk and the shear moduli of microgel suspensions
originates from the high compressibility of a microgel. This makes the geometric features of these suspensions concentration independent,
and their mechanical response characterized by a high and concentration independent degree of non-affinity.
However, a direct measure of the non-affine response of suspensions of compressible particles is still lacking,
and we hope that this work will stimulate research in this direction.

\section{Acknowledgements}
We thank A. Coniglio and A. Fern\'{a}ndez-Nieves for their helpful comments on an earlier version of the manuscript.
MPC acknowledges financial support from MIUR-FIRB RBFR081IUK.
%\footnotetext{\dag~Electronic Supplementary Information (ESI) available: [details of any supplementary information available should be included here]. See DOI: 10.1039/b000000x/}
%Please use \dag to cite the ESI in the main text of the article.
%If you article does not have ESI please remove the the \dag symbol from the title and the above footnotetext.

\footnotetext{\textit{$^{a}$~Italian Institute of Technology @ CRIB, Napoli, Italy; E-mail: giovanni.romeo@iit.it}}
\footnotetext{\textit{$^{b}$~Department of Materials Engineering and Production, University of Napoli Federico II, Italy; }}
\footnotetext{\textit{$^{c}$~CNR--SPIN, Dipartimento di Scienze Fisiche, University of Napoli Federico II, Italy}}

%additional addresses can be cited as above using the lower-case letters, c, d, e... If all authors are from the same address, no letter is required

%\footnotetext{\ddag~Additional footnotes to the title and authors can be included \emph{e.g.}\ `Present address:' or `These authors contributed equally to this work' as above using the symbols: \ddag, \textsection, and \P. Please place the appropriate symbol next to the author's name and include a \texttt{\textbackslash footnotetext} entry in the the correct place in the list.}

\footnotesize{
\bibliographystyle{rsc}

}

\end{document}